\documentclass[table]{ccjnl}
\pdfoutput=1
\title{AoI-Driven Queue Management and Power Control in V2V Networks: A GNN-Enhanced MARL Approach}
\author{
Hao Fang \inst{1}, Xiao Li \inst{1,*}, Chongtao Guo \inst{2}, Le Liang \inst{1,*}, Shi Jin \inst{1}
\corinfo{$\textrm{li\_xiao}$@seu.edu.cn; $\textrm{lliang}$@seu.edu.cn}}

\receiveddate{April 4, 2023}
\reviseddate{July 10, 2023}
\Editor{Honggang Zhang}

\address[1]{The National Mobile Communications Research Laboratory, Southeast University, Nanjing 210096, China}
\address[2]{The Guangdong Key Laboratory of Intelligent Information Processing, College of Electronics and Information Engineering, Shenzhen University, Shenzhen 518060, China}

\begin{document}
\maketitle

\begin{abstract}
Queue management and resource allocation play a critical role in enabling cooperative status awareness in vehicular networks. This paper investigates the problem of age of information (AoI)-aware status updates in vehicle-to-vehicle (V2V) communication, where each vehicle's status is represented by multiple interdependent packets. To enable fine-grained queue management at the packet level under resource constraints, we formulate a joint optimization problem that simultaneously learns active packet dropping and transmit power control strategies. A hybrid action space is designed to support both discrete dropping decisions and continuous power control. To exploit the graph-structured interference inherent in V2V topology, a graph neural network (GNN) is introduced to aggregate slowly varying large-scale fading, allowing agents to capture topological dependencies implicitly without frequent message exchange. The overall framework is built upon multi-agent proximal policy optimization (MAPPO), with centralized training and decentralized execution (CTDE). Simulations demonstrate that the proposed method significantly reduces average AoI across a wide range of network densities, channel conditions, and traffic loads, consistently outperforming several baselines.
\end{abstract}

\section{introduction}\label{s1}
The advancement of autonomous driving has brought vehicle intelligence into a new phase, where the role of vehicle-to-everything (V2X) communication is becoming increasingly central \cite{1, 39}. While onboard sensing and decision-making remain fundamental, they are no longer sufficient to ensure safety and efficiency in complex and dynamic traffic environments. Fortunately, vehicle-to-vehicle (V2V) communication can enable the timely exchange of status and intent information among vehicles, facilitating cooperative awareness and coordinated decision-making \cite{2}. By extending each vehicle's perceptual range beyond line-of-sight and enabling low-latency interaction with both surrounding vehicles and roadside infrastructure, V2X functions not merely as a communication medium, but as a foundational enabler of collective intelligence in autonomous systems operating under uncertainty \cite{3, 40}.

Unlike traditional autonomous systems that operate in isolation, cooperative driving relies on continuous negotiation and synchronization among multiple vehicles to enable safety-critical functionalities, such as coordinated lane merging, intersection crossing, and collision avoidance in obstructed or congested environments \cite{4}. These cooperative behaviors hinge on the timely and reliable dissemination of high-dimensional status information, including vehicle position, velocity, acceleration, and driving intent. The timely exchange of such information enhances the stability and responsiveness of distributed control decisions \cite{5, 6}. However, traditional metrics such as throughput, latency, and reliability, are less effective to evaluate the performance of such a status update communication system \cite{7}.

Age of information (AoI) \cite{20, 38}, defined as the time elapsed since the generation of the latest received packet, quantifies the freshness of the transmitter's status information from the receiver's perspective. Maintaining a low AoI enables the receiver to make decisions based on fresher environment awareness, which is critical for ensuring safety and effective coordination in distributed systems. Given the highly dynamic nature of vehicular networks, where channel state information (CSI) and topology change frequently, timely status updates are critical to ensuring real-time awareness and reliable cooperative behavior.

Effective resource allocation is essential for AoI reduction, as it directly affects the number of status update packets that can be transmitted over a constrained wireless channel. Traditional model-based optimization techniques, such as linear programming, integer programming, and convex optimization, have been extensively studied for resource management \cite{10, add10, 23}. To ensure that extreme AoI violations remain within acceptable bounds under power and latency constraints, a tail-optimal AoI control framework was proposed in \cite{19}, which directly minimizes the probabilistic AoI tail distribution through joint transmission power and delay optimization. AoI-aware resource allocation has also been explored in vehicular networks, where a closed-form expression of the AoI outage probability for D/Geo/1 queueing systems was derived to characterize V2V status update performance \cite{18}. However, these approaches typically assume full knowledge of system parameters, including instantaneous CSI, packet arrival statistics, and network topology. 

To overcome the limitations of static model-based approaches, reinforcement learning (RL) has emerged as a promising data-driven solution that can adapt to dynamic and time-varying wireless environments \cite{11}. A representative study \cite{14} proposed a multi-agent reinforcement learning (MARL)-based distributed spectrum sharing scheme for vehicular networks, where V2V agents learn to coordinate spectrum access and power control through fingerprint-enhanced deep Q-networks. The work in \cite{15} tackled power control in generic wireless networks with a focus on robustness to CSI delay and large-scale deployment. Age-aware radio resource management has been explored using deep reinforcement learning, where a proactive scheduling policy is learned to minimize the long-term average AoI of vehicular links under dynamic channel conditions and traffic patterns \cite{22}. A multi-agent multi-task reinforcement learning framework has been developed to jointly optimize spectrum reuse and transmit power in \cite{24}, aiming to minimize the average AoI across all V2V links while satisfying ultra-reliable and low-latency communication constraints. However, RL models lack the structural inductive ability to capture dynamic topological dependencies, and thus fail to exploit the spatial correlations essential for scalable coordination and efficient learning.



Graph reinforcement learning (GRL) \cite{25, 26, 41} offers a principled framework for capturing the dynamic and structured characteristics of vehicular communication networks. Vehicular networks are naturally graph-structured systems \cite{17}, where the communication topology evolves rapidly due to vehicle mobility. By integrating graph neural networks (GNNs) as function approximators within the RL architecture, GRL enables the extraction of rich representations that encapsulate both local interactions and global network states. The study in \cite{27} tackled AoI-aware spectrum sharing via a distributed actor-critic framework, where graph attention networks were integrated into a global edge-assisted critic to enable scalable and topology-aware policy evaluation under uncertainty. A GRL-based framework was proposed in \cite{28} to minimize user-side AoI by enabling scalable and topology-aware UAV coordination under partial observability, through the integration of GNNs and the QMIX algorithm within a centralized training and distributed execution (CTDE) paradigm. Despite these advances, most existing GRL-based approaches for AoI optimization tend to overlook the communication and coordination overhead required to maintain global or near-global network awareness. This issue becomes particularly critical in AoI-sensitive scenarios where timeliness is of paramount importance.

Furthermore, most existing studies on cooperative driving assume that each vehicle's status can be encapsulated into a single data packet for sharing with neighboring vehicles \cite{29, 30, 31, 32}. Although this assumption simplifies system modeling, it does not capture the high-dimensional and multi-modal nature of real-world collaborative perception data. In practical perception pipelines, a vehicle's status is represented by rich sensory observations such as LiDAR point clouds, multi-camera feature maps, and radar occupancy grids. These data entities consist of tens of thousands of points or high-resolution feature tensors and therefore cannot be transmitted within a single packet; instead, they must be fragmented into multiple packets prior to transmission. Consequently, the freshness and utility of a received perception status depend on the successful delivery of a sufficiently large subset of its packet sequence, rendering conventional single-packet AoI models inadequate for collaborative perception. The batch-based status representation introduced in prior work \cite{37} partially addresses this limitation by modeling each status update as a sequence of packets and optimizing power allocation and sampling under AoI constraints. However, the framework in \cite{37} is restricted to a centrally coordinated setting with periodically generated batches and overlooks the fact that the utility of a received perception status depends on the successful delivery of a sufficiently complete subset of its constituent packets. In addition, it does not incorporate decentralized decision-making, packet-level queue management, or topology-aware interference modeling, all of which are essential for highly dynamic vehicular environments.

In this work, we extend the batch-based status-update framework to a more realistic collaborative perception setting by jointly learning active packet-dropping and transmit power control policies through a multi-agent reinforcement learning (MARL) approach. The objective is to minimize network-wide AoI while incorporating resource limitations, decentralized operation, and the graph-structured characteristics of vehicular interference. The main contributions of this paper are summarized as follows.
\begin{itemize}
\item We propose a fine-grained AoI optimization framework that jointly learns packet-level queue management and transmit power control through a hybrid discrete-continuous action space. This design is well suited for collaborative perception, where multi-packet sensory data must be delivered under stringent latency and resource constraints.
\item We adopt GraphSAGE to model the interference topology in vehicular networks. By aggregating large-scale fading information at a coarse timescale, this approach achieves a balance between topological representation capability and communication overhead.
\item We develop a CTDE framework based on multi-agent proximal policy optimization (MAPPO) \cite{36} with an AoI-centric reward, allowing decentralized agents to learn scalable coordination strategies. Extensive experiments demonstrate that our method consistently outperforms baselines across diverse scenarios, including varying packet sizes, traffic loads, and channel conditions.
\end{itemize}

The remainder of this paper is organized as follows. We introduce the system model and formulate the problem in Section \ref{s2}. In Section \ref{s3}, the GNN-enhanced CTDE framework is described in detail, including its key components and algorithmic structure. Simulation results are presented in Section \ref{s4}. Finally, the conclusion is provided in Section \ref{s5}.

\begin{figure*}[t]
\begin{center}
\vspace{-3mm}
  \includegraphics[width=6.6in,height=1.8in]{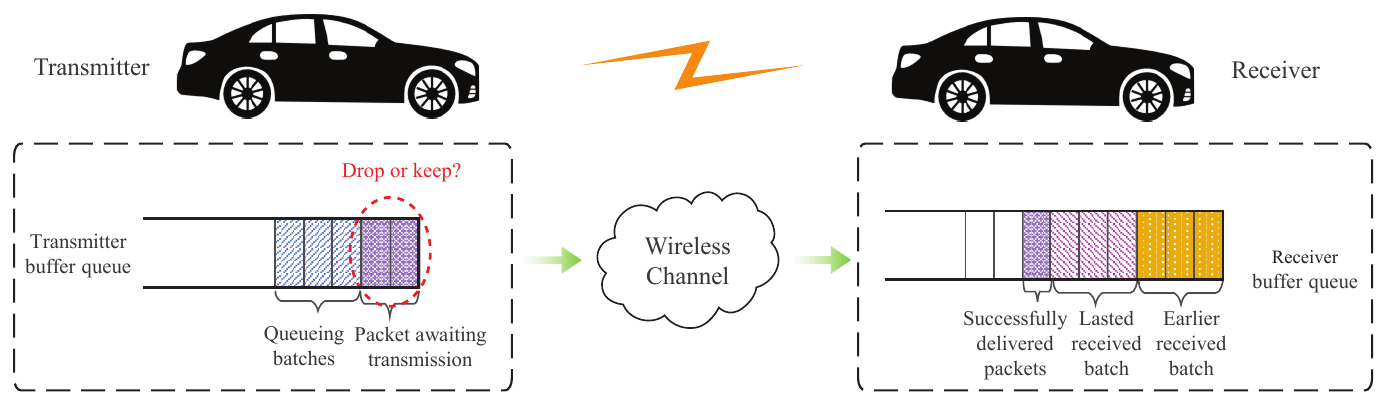}
  \caption{Illustration of queueing model in a V2V communication scenario with batch size $u = 3$.}\label{system_model}
\end{center}
\vspace{-20pt}
\end{figure*}

\section{System Model and Problem Formulation}\label{s2}
In this section, we introduce the system model and formulate the AoI-aware queue management and power control problem under a multi-packet vehicle status representation. Time is discretized into slots of fixed and small duration to match the fine-grained temporal resolution required for low-latency communication in vehicular networks. In the wireless environment, both large-scale and small-scale fading are considered, but they evolve at different temporal rates. Large-scale fading, which includes effects such as path loss and shadowing, changes slowly over time and is assumed to remain constant within each interval of $N$ slots. Therefore, only once channel information collection in every $N$ slots can be enough for the system to capture the network topology. In contrast, small-scale fading varies rapidly due to multipath propagation and vehicle mobility and is updated in every slot. These real-world channel characteristics allow long-term topology-aware representation while allowing adaptation to short-term channel variations, which is essential for scalable and adaptive resource management.

\subsection{System Model}
Consider a vehicular network with $M$ V2V communication links, where each link is responsible for transmitting status updates from a dedicated transmitter to its corresponding receiver. All links operate over a shared frequency band, resulting in mutual interference among simultaneous transmissions. To ensure the timeliness of status dissemination, power control must be intelligently coordinated to mitigate interference, while dynamically managing the queues of fragmented status packets. Let ${\cal M} = \left\{ {1, \cdots ,M} \right\}$ denote the set of V2V communication links, and ${\cal N} = \left\{ {1, \cdots ,N} \right\}$ represent the set of time slots. Let ${h_{mm}}\left[ n \right]$ denote the direct channel coefficient of the $m$-th V2V link at time slot $n$, and ${h_{im}}\left[ n \right]$ denote the interference channel coefficient from the $i$-th transmitter to the $m$-th receiver at time slot $n$, where $i\ne m$. The achievable channel capacity of the $m$-th V2V link in time slot $n$ is given by
\begin{equation}\label{e1}
{C_m}\left[ n \right] \!= \! B{\log _2}\!\left(\! {1 \!+\! \frac{{{{\left| {{h_{mm}}\left[ n \right]} \right|}^2}{p_m}\left[ n \right]}}{{{\sigma ^2} + \sum\limits_{i = 1,i \ne m}^M {{{\left| {{h_{im}}\left[ n \right]} \right|}^2}{p_i}\left[ n \right]} }}} \!\right),
\end{equation}
where ${p_m}\left[ n \right]$ denotes the transmit power of the $m$-th V2V link in the slot $n$, for all $m \in {\cal M}$ and $n \in {\cal N}$, $\sigma^2$ is the noise variance, and $B$ is the bandwidth. Theoretically, the number of packets that can be successfully transmitted over the $m$-th V2V link during time slot $n$ is given by
\begin{equation}\label{e2}
{y_m}\left[ n \right] =  \left\lfloor {\frac{{t \cdot {C_m}\left[ n \right]}}{L}} \right\rfloor ,
\end{equation}
where $t$ denotes the duration of a time slot, $L$ denotes the size of a single packet in bits, and the floor operation, $\left\lfloor  \cdot  \right\rfloor $, ensures that only fully transmittable packets are counted, reflecting the discrete nature of data transmission.

Since a single status may require multiple packets for complete representation, as illustrated in Fig. \ref{system_model}, data sampling at the vehicle transmitter is performed in batches, each consisting of $u \in {\mathbb{N}^+ }$ packets. In each time slot, a new batch arrives with a certain probability, independently across time. Let ${\rho _m}\left[ n \right] \in \left\{ {0,1} \right\}$ denote the batch arrival indicator for transmitter $m$ at time slot $n$, where ${\rho _m}\left[ n \right] = 1$ if a new batch arrives during the slot $n$, and ${\rho _m}\left[ n \right] = 0$ otherwise.

\begin{figure*}[t]
\begin{center}
  \includegraphics[width=5.8in,height=4.2in]{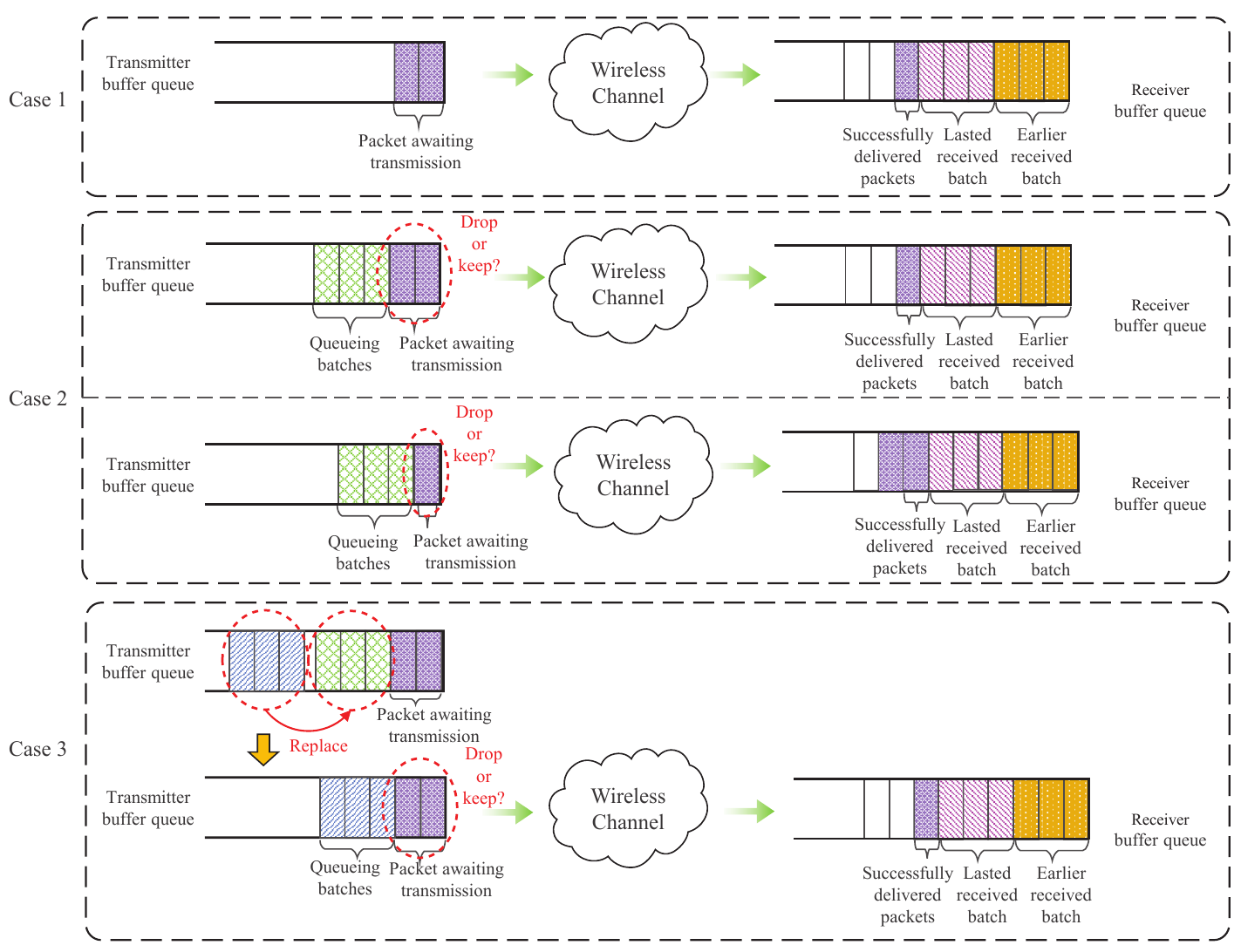}
  \caption{ Illustration of queue management with batch size $u = 3$, including packet replacement and dropping rules.}\label{buffer_case}
\end{center}
\end{figure*}

Since the receiver can reconstruct the original status representation only when all packets within a batch are successfully received, a critical problem arises in the presence of partial transmissions. As illustrated in Fig. \ref{buffer_case}, consider the case where the transmission of a current status representation has not yet been completed, while a new batch of data packets corresponding to the next status has already arrived. This scenario poses a fundamental scheduling dilemma: whether to continue transmitting the remaining packets of the current batch or to drop the incomplete batch and switch to transmitting the newly arrived data. These two strategies lead to different impacts on the AoI, which necessitates a thorough investigation. To address this trade-off, a packet dropping factor ${\gamma _m}\left[ n \right] \in \left\{ {0,1} \right\}$ is introduced. Specifically, ${\gamma _m}\left[ n \right] = 0$ indicates that the remaining untransmitted packets of the current batch are dropped in favor of the newly arrived batch, whereas ${\gamma _m}\left[ n \right] = 1$ implies that the transmission of the current batch continues without interruption.

As illustrated in Fig. \ref{buffer_case}, case 1 corresponds to a scenario where only a single batch of data packets resides in the buffer. In this case, proactive packet dropping is prohibited, and the ongoing transmission continues until completion. In contrast, case 3 describes a more complex situation in which the current batch remains only partially transmitted while multiple new batches have already arrived. In this case, the next batch in line, which was originally scheduled to be transmitted after the current batch, is replaced by the most recently arrived batch. It is under this condition that the packet dropping factor ${\gamma _m}\left[ n \right]$ is reconsidered to determine whether to continue with the current batch or to drop it in favor of the newly arrived data. As a result, the queue length is bounded by two batches, and the buffer only needs to accommodate at most two batches of data packets. Let ${q_m}\left[ n \right] = \left( {q_m^{\left( 1 \right)}\left[ n \right],q_m^{\left( 2 \right)}\left[ n \right]} \right)$ denote the queue length of transmitter $m$ at the beginning of time slot $n$, where $q_m^{\left( 1 \right)}\left[ n \right]$ represents the number of remaining packets in the earlier-arrived batch, and $q_m^{\left( 2 \right)}\left[ n \right]$ corresponds to the number of remaining packets in the later-arrived batch. The age of these two batches at the transmitter side is denoted by $\Delta _m^{\texttt{Tr}}\left[ n \right] = \left( {\Delta _m^{\left( 1 \right)}\left[ n \right],\Delta _m^{\left( 2 \right)}\left[ n \right]} \right)$, where $\Delta_m^{(i)}[n]$ represents the time having elapsed since the correponding batch entered the queue for $i=1, 2$. Let $\Delta _m^{\texttt{Re}}\left[ n \right]$ denote the AoI at the receiver of the $m$-th V2V link at time slot $n$, which reflects the freshness of successfully received data.

\subsection{Problem Formulation}
At the beginning of each time slot, the queue length, the age at the transmitter side, and the AoI at the receiver are updated based on the current content and newly arriving data. The updated rules can be summarized as follows.

\textbf{\subsubsection{Queue Management}}
At the beginning of each time slot, the buffer executes a packet replacement and dropping process according to the current queue status, as illustrated in Fig. \ref{buffer_case}.

Case 1: The buffer is empty, i.e., $q_m^{\left( 1 \right)}\left[ n \right] = q_m^{\left( 2 \right)}\left[ n \right] = 0$. In this case, if a new batch arrives, it is directly stored in the buffer as the earlier batch, with its age initialized to zero. In this case, no dropping or replacement is needed, and the queue length and the age at the transmitter side are represented as
\begin{equation}\label{e3}
q_m[n] =
\begin{cases}
(0, 0),        & \!\! \text{if } \rho_m[n] = 0, \\
(u, 0),        & \!\! \text{if } \rho_m[n] = 1.
\end{cases}
\end{equation}

\begin{equation}\label{e4}
\Delta_m^{\text{Tr}}[n] =
\begin{cases}
(\infty, \infty), & \!\! \text{if } \rho_m[n] = 0, \\
(0, \infty),      & \!\! \text{if } \rho_m[n] = 1.
\end{cases}
\end{equation}
where $\infty$ indicates that no packet is present in the current time slot.

Case 2: One batch of packets already exists in the buffer, i.e., $q_m^{\left( 1 \right)}\left[ n \right] > 0$ and $q_m^{\left( 2 \right)}\left[ n \right] = 0$. If a new batch arrives at the beginning of the time slot, it is stored in the buffer as the later batch with an initial age of zero. No replacement occurs, and both batches are retained. The update process of the queue length and the age at the transmitter side are as follows.
\begin{equation}\label{e5}
q_m[n] =
\begin{cases}
\left( q_m^{(1)}[n],\, 0 \right), & \!\! \text{if } \rho_m[n] = 0, \\
\left( u,\, 0 \right),            & \!\! \text{if } \rho_m[n] = 1, \gamma_m[n] = 0, \\
\left( q_m^{(1)}[n],\, u \right), & \!\! \text{if } \rho_m[n] = 1, \gamma_m[n] = 1.
\end{cases}
\end{equation}
\begin{equation}\label{e6}
\Delta_m^{\text{Tr}}[n] \!=\!
\begin{cases}
\left( \Delta_m^{(1)}[n],\, \infty \right), & \!\! \text{if } \rho_m[n] = 0, \\
\left( 0,\, \infty \right),                & \!\! \text{if } \rho_m[n] = 1, \gamma_m[n] = 0, \\
\left( \Delta_m^{(1)}[n],\, 0 \right),     & \!\! \text{if } \rho_m[n] = 1, \gamma_m[n] = 1.
\end{cases}
\end{equation}

Case 3: Two batches of data packets are already present in the buffer, i.e., $q_m^{\left( 1 \right)}\left[ n \right] > 0$ and $q_m^{\left( 2 \right)}\left[ n \right] > 0$. If a new batch arrives at the beginning of the time slot, the later-arrived batch (i.e., the second batch) in the buffer is replaced by the newly arrived one. The age of the freshly arrived batch is initialized to zero. At this point, the system must decide whether to continue transmitting the earlier batch or to drop it based on the packet dropping factor. The queue length and the age at the transmitter side are then updated accordingly
\begin{equation}\label{e7}
q_m[n] \!=\! \left\{\!\!\!
\begin{array}{ll}
\left( q_m^{(2)}[n],\, 0 \right),                      & \!\!\! \text{if } \rho_m[n] \!=\! 0,\!\; \gamma_m[n] \!=\! 0, \\
\left( q_m^{(1)}[n],\, q_m^{(2)}[n] \right),           & \!\!\! \text{if } \rho_m[n] \!=\! 0,\!\; \gamma_m[n] \!=\! 1, \\
\left( u,\, 0 \right),                                 & \!\!\! \text{if } \rho_m[n] \!=\! 1,\!\; \gamma_m[n] \!=\! 0, \\
\left( q_m^{(1)}[n],\, u \right),                      & \!\!\! \text{if } \rho_m[n] \!=\! 1,\!\; \gamma_m[n] \!=\! 1.
\end{array}
\right.
\end{equation}

\begin{equation}\label{e8}
\Delta_m^{\text{Tr}}[n] \!\!=\!\! \left\{\!\!\!\!
\begin{array}{ll}
\left( \Delta_m^{(2)}[n],\,\! \infty \right)\!,                        & \!\!\!\! \text{if } \rho_m[n] \!=\! 0,\!\;\! \gamma_m[n] \!=\! 0, \\
\left( \Delta_m^{(1)}[n],\,\! \Delta_m^{(2)}[n] \right)\!,             & \!\!\!\! \text{if } \rho_m[n] \!=\! 0,\!\;\! \gamma_m[n] \!=\! 1, \\
\left( 0,\,\! \infty \right)\!,                                        & \!\!\!\! \text{if } \rho_m[n] \!=\! 1,\!\;\! \gamma_m[n] \!=\! 0, \\
\left( \Delta_m^{(1)}[n],\,\! 0 \right)\!,                             & \!\!\!\! \text{if } \rho_m[n] \!=\! 1,\!\;\! \gamma_m[n] \!=\! 1.
\end{array}
\right.
\end{equation}

\textbf{\subsubsection{Resource Allocation}}
After each time slot, the buffer is updated according to the transmission outcomes. A first-come-first-served (FCFS) policy is adopted, whereby packets from the earlier-arrived batch, $q_m^{\left( 1 \right)}\left[ n \right]$, are prioritized for transmission. If the transmission capacity in the current time slot exceeds the size of the earlier batch, the remaining capacity is allocated to packets in the later-arrived batch, $q_m^{\left( 2 \right)}\left[ n \right]$.

\begin{itemize}
\item Partial transmission of the earlier batch:

If ${y_m}\left[ n \right] < q_m^{\left( 1 \right)}\left[ n \right]$, then only a portion of the earlier batch is transmitted. In this case, the transmission capacity is fully consumed by the earlier batch, and the later-arrived batch remains untouched. The queue length, the age at the transmitter side, and the AoI at the receiver are updated as
\begin{equation}\label{e9}
{q_m}\left[ {n + 1} \right] = \left( {q_m^{\left( 1 \right)}\left[ n \right] - {y_m}\left[ n \right],q_m^{\left( 2 \right)}\left[ n \right]} \right),
\end{equation}
\begin{equation}\label{e10}
\Delta _m^{{\rm{Tr}}}\left[ {n + 1} \right] = \left( {\Delta _m^{\left( 1 \right)}\left[ n \right] + 1,\Delta _m^{\left( 2 \right)}\left[ n \right] + 1} \right),
\end{equation}
\begin{equation}\label{e11}
\Delta _m^{{\mathop{\rm Re}\nolimits} }\left[ {n + 1} \right] = \Delta _m^{{\mathop{\rm Re}\nolimits} }\left[ n \right] + 1.
\end{equation}

\item Full transmission of the earlier batch and partial transmission of the later batch:

If the transmission capacity in time slot $n$ is sufficient to fully transmit the earlier-arrived batch and partially transmit the later-arrived batch, $q_m^{\left( 1 \right)}\left[ n \right] \le {y_m}\left[ n \right] < q_m^{\left( 1 \right)}\left[ n \right] + q_m^{\left( 2 \right)}\left[ n \right]$, all
$q_m^{\left( 1 \right)}\left[ n \right]$ packets in the first batch are transmitted, and the remaining capacity is allocated to the second batch. The queue length, the age at the transmitter side, and the AoI at the receiver are updated as
\begin{equation}\label{e12}
\begin{array}{l}
{q_m}[n + 1] = \\
~~~~~~~~\left( {\max \left( {q_m^{(2)}[n] - {y_m}[n] + q_m^{(1)}[n],0} \right),0} \right),
\end{array}
\end{equation}

\begin{equation}\label{e13}
\Delta _m^{{\rm{Tr}}}\left[ {n + 1} \right] = \left( {\Delta _m^{\left( 2 \right)}\left[ n \right] + 1,\infty } \right),
\end{equation}
\begin{equation}\label{e14}
\Delta _m^{{\mathop{\rm Re}\nolimits} }\left[ {n + 1} \right] = \Delta _m^{\left( 1 \right)}\left[ n \right] + 1.
\end{equation}

\item Full transmission of both batches:

If the available transmission capacity in time slot $n$ is sufficient to transmit all remaining packets in both the earlier and later-arrived batches, ${y_m}\left[ n \right] \ge q_m^{\left( 1 \right)}\left[ n \right] + q_m^{\left( 2 \right)}\left[ n \right]$, As a result, the buffer is completely emptied at the end of the time slot. The queue length, the age at the transmitter side, and the AoI at the receiver are updated as
\begin{equation}\label{e15}
{q_m}\left[ {n + 1} \right] = \left( {0,0} \right),
\end{equation}
\begin{equation}\label{e16}
\Delta _m^{{\rm{Tr}}}\left[ {n + 1} \right] = \left( {\infty ,\infty } \right),
\end{equation}
\begin{equation}\label{e17}
\Delta _m^{{\mathop{\rm Re}\nolimits} }\left[ {n + 1} \right] = \Delta _m^{\left( 2 \right)}\left[ n \right] + 1.
\end{equation}

\end{itemize}

\textbf{\subsubsection{Problem Formulation}}
It is worth noting that the best decision of transmitting or dropping a data packet within a given time slot is inherently affected by the transmit power level. Consequently, it is essential to jointly optimize the transmit power $\textbf{p}\left[ n \right]$ and the packet dropping factor $\boldsymbol{\mit\gamma} \left[ n \right]$ to minimize the long-term average AoI. The corresponding optimization problem is formulated as
\begin{equation}\label{e18}
\begin{array}{l}
\mathop {\min }\limits_{{{\left\{ {\textbf{p}\left[ n \right],\boldsymbol{\mit\gamma} \left[ n \right]} \right\}}_{n \in {\cal N}}}} {\rm{   }}\frac{1}{{MN}}\sum\limits_{n = 1}^N {\sum\limits_{m = 1}^M {\Delta _m^{{\mathop{\rm Re}\nolimits} }\left[ n \right]} } ,\\
{\rm{s}}{\rm{.t}}{\rm{.~~~~0}} \le {p_m}\left[ n \right] \le {P_{\max }},\forall m \in {\cal M},n \in {\cal N},\\
{\rm{~~~~~~~~~}}{\gamma _m}\left[ n \right] \in \left\{ {0,1} \right\},\forall m \in {\cal M},n \in {\cal N},
\end{array}
\end{equation}
where $\textbf{p}\left[ n \right] = {\left[ {{p_1}\left[ n \right], \cdots ,{p_M}\left[ n \right]} \right]^T}$ and $\boldsymbol{\mit\gamma}\left[ n \right] = {\left[ {{\gamma _1}\left[ n \right], \cdots ,{\gamma _M}\left[ n \right]} \right]^T}$ denote the transmit power vector and the packet dropping factor vector, respectively. Each transmit power is constrained by a maximum power value ${P_{\max }}$.

\section{The GNN-enhanced CTDE framework}\label{s3}

\begin{figure*}[t]
\begin{center}
\vspace{-3mm}
  \includegraphics[width=6.1in,height=2.4in]{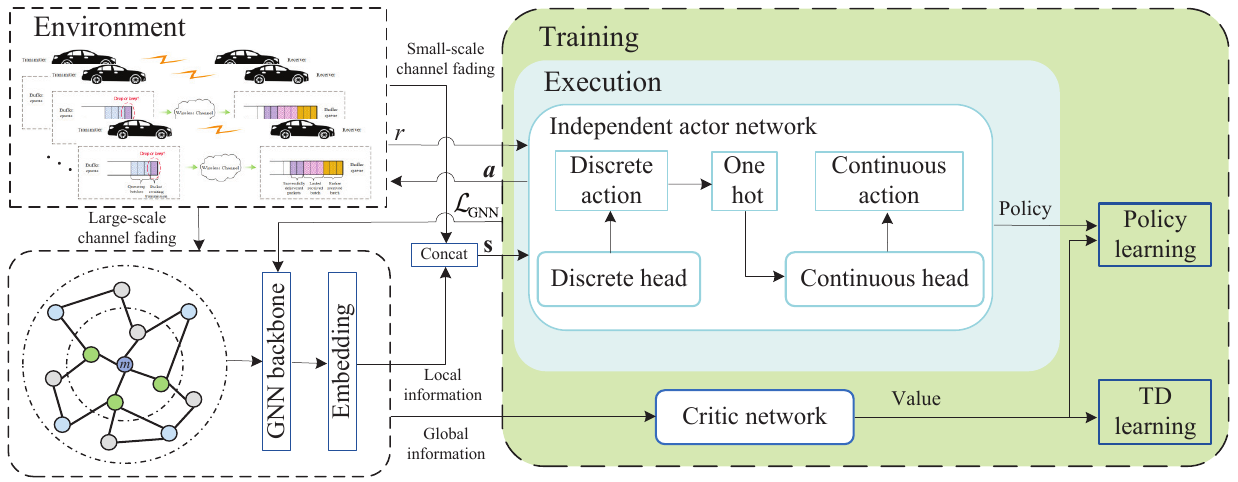}
  \caption{The proposed GNN-enhanced CTDE framework based on MAPPO.}\label{algorithm_architecture}
\end{center}
\vspace{-15pt}
\end{figure*}

The joint design of packet dropping and transmit power control constitutes a sequential decision-making problem, as each action affects future queue dynamics, AoI evolution, and interference levels. Since transmission performance is heavily shaped by topology-dependent mutual interference, a scalable representation of link interactions is necessary. GNNs provide an effective means to capture this interference structure, and their integration with MARL enables decentralized agents to make long-term AoI-aware decisions. These considerations motivate the GNN-enhanced MARL framework illustrated in Fig. \ref{algorithm_architecture}.
%

\subsection{GNN-based Graph Representation}
In cooperative autonomous driving scenarios, the wireless communication network formed by V2V links exhibits inherent spatial structure and dynamic topology. Traditional flat-status representations fail to capture the complex interference and interdependence among the V2V links in the network. To address this limitation, we model the communication environment as a graph and utilize a GNN to learn topology-aware status embeddings for informed decision-making.

We represent the V2V communication network as a directed graph $\mathcal{G} = (\mathcal{V}, \mathcal{E})$, where $\mathcal{V}$ and $\mathcal{E}$ are the set of nodes and edges, repectively. Each node ${v_m} \in \mathcal{V}$ corresponds to a V2V link. Directed edge ${e_{mj}} \in \mathcal{E}$ represents potential interference relationships from link $j$ to link $m$. Each node is described by a node feature vector $\mathbf{x}_m \in {\mathbb{R}^5}$:
\begin{equation}\label{e19}
    \mathbf{x}_m = \left[ \rm{PL}_m,\, p^{\rm{tx}}_{x,m},\, p^{\rm{tx}}_{y,m},\, p^{\rm{rx}}_{x,m},\, p^{\rm{rx}}_{y,m} \right],
\end{equation}
where $\rm{PL}_m$ denotes the path loss of the $m$-th V2V link, $\rm{p_{x,m}^{tx}}$ and $\rm{p_{y,m}^{tx}}$ represent the horizontal and vertical coordinates of the transmitter, respectively, and $\rm{p_{x,m}^{rx}}$ and $\rm{p_{y,m}^{rx}}$ represent those of the receiver.

To efficiently capture spatial correlations and interference patterns, we adopt a two-layer GraphSAGE network with neighborhood sampling to enhance scalability by limiting the number of neighbors involved in each aggregation. This property is particularly valuable in AoI-sensitive vehicular networks, where dense connectivity can impose significant communication overhead. To further reduce communication overhead and avoid excessive information exchange, we construct the graph using only large-scale channel fading, which evolves slowly and is updated once every $N$ slots rather than at every time slot. This design alleviates the overhead of frequent GNN aggregation while improving the stability of the extracted topological features.

Unlike traditional GraphSAGE implementations with a fixed sample size, our model dynamically determines the number of sampled neighbors for each node based on the current graph size. Specifically, for a graph with $M$ nodes, the number of sampled neighbors, denoted by $\mathcal{D}$, for each node is given by
\begin{equation}
 \mathcal{D} = \max \left(2, \min(M - 1,\, \lfloor 2 \sqrt{M} \rfloor ) \right).
\end{equation}

This allows the GNN to adaptively balance between representation richness and computational efficiency across varying network sizes and densities. Each layer performs neighborhood aggregation and feature transformation as follows
\begin{equation}
    \boldsymbol{\Omega}^{(k+1)}_m \!= \!\sigma\!\left( \mathbf{W}_k \left[ \boldsymbol{\Omega}^{(k)}_m \mathbin{\|} \frac{1}{|\mathcal{D}_m^{(k)}|} \sum_{j \in \mathcal{D}_m^{(k)}} \boldsymbol{\Omega}^{(k)}_j \right] \right),
\end{equation}
where $\boldsymbol{\Omega}^{(k)}_m$ is the hidden representation of node $m$ at layer $k$, $\sigma(\cdot)$ is the activation function, and $\mathcal{D}_m^{(k)}$ denotes the sampled neighborhood at layer $k$. The symbol $\mathbin{\|}$ represents a concatenation operation along the feature dimension, enabling the model to jointly encode self and neighbor information before the transformation.

After two rounds of aggregation, the final output embedding of node $m$ is denoted by
\begin{equation}
    f_m = \boldsymbol{\Omega}^{(2)}_m \in \mathbb{R}.
\end{equation}
This scalar-valued embedding $f_m$ is subsequently used as part of the agent's local state input for decision-making. We employ two GraphSAGE layers so that each link can aggregate information from its two-hop interference neighborhood, which captures the main sources of mutual interference in V2V networks. Deeper GNNs offer limited gains while increasing computational cost and causing over-smoothing. Thus, a two-layer encoder provides a good tradeoff between accuracy and efficiency.

\subsection{RL Formulation}
The proposed problem in V2V networks can be naturally formulated as a Markov decision process (MDP), where each V2V link operates as a learning agent interacting with a dynamic wireless environment. At each time slot, the agent observes its local state, selects a joint transmission action, and receives a feedback reward based on the resulting AoI performance. In the following subsections, we detail the components of the RL formulation, including the state representation, action space, and reward design.

\textbf{\subsubsection{State Representation}}
In the proposed framework, each V2V link is modeled as an autonomous agent. To enable decentralized decision-making while minimizing system-wide average AoI, it is critical for each agent to maintain an expressive local-state representation. In realistic vehicular scenarios, the generation and delivery of high-fidelity status information are subject to both structural and operational constraints.

Specifically, a complete status update may span multiple packets due to its high dimensionality, while the underlying data generation process is inherently stochastic and occurs in batches. This design must reflect not only the physical-layer channel dynamics but also the freshness and urgency of buffered information. Due to these considerations, the state $\mathbf{s}_m \in \mathbb{R}^7$ for agent $m$ at slot $n$ is defined as
\begin{equation}
\begin{array}{c}
{{\bf{s}}_m}[n] = \left[ {{f_m},{\mkern 1mu} h_m^{{\rm{fast}}}[n],{\mkern 1mu} q_m^{\left( 1 \right)}[n]} \right.,\Delta _m^{\left( 1 \right)}[n],\\
\left. ~~~~~~~~~~~~~~~~~~~~~~~~~~~~{q_m^{\left( 2 \right)}[n],{\mkern 1mu} \Delta _m^{\left( 2 \right)}[n],{\mkern 1mu} \Delta _m^{{\rm{Re}}}[n]} \right],
\end{array}
\end{equation}
where $f_m \in \mathbb{R}$ is the GNN-generated feature embedding for link $m$, derived from large-scale channel fading and shared among all agents at the beginning of each episode. Here, an episode refers to a decision-making horizon comprising $N$ consecutive time slots, during which the large-scale channel fading is assumed to remain constant. The term $h_m^{\text{fast}}[n]$ denotes the instantaneous small-scale fading coefficient of link $m$ in slot $n$, incorporated into each agent's local state to capture rapid channel fluctuations. The combination of the slowly varying embedding $f_m $ and the per-slot fading term $h_m^{\text{fast}}[n]$ yields a two-timescale representation that captures long-term topology characteristics while enabling responsiveness to fast fading.

\textbf{\subsubsection{Action Space}}
Each agent in the proposed framework operates within a hybrid action space that combines a discrete packet-dropping decision with a continuous power allocation. Specifically, at slot $n$, agent $m$ selects an action $a_m[n] = (\gamma _m[n], p_m[n])$. The discrete decision enables freshness-aware queue control, allowing agents to drop stale or unproductive batches. In contrast, the continuous decision coordinates physical-layer interference, as the data rate is directly influenced by the transmit power and channel fading.

Formally, the action space $\mathcal{A}_m$ for agent $m$ is defined as
\begin{equation}
\mathcal{A}_m = \left\{ {0,1} \right\} \times [0, P_{\max}],
\end{equation}
where each action consists of a binary packet-dropping decision in $\left\{ {0,1} \right\}$ and a continuous transmit-power level in $[0, P_{\max}]$, forming a hybrid discrete-continuous action space through the Cartesian product. And the joint action space is $\mathcal{A} = \prod_{m=1}^{M} \mathcal{A}_m$.

\textbf{\subsubsection{Reward Design}}
The objective of the RL framework is to minimize the long-term average AoI across all V2V links. To align the per-agent learning process with this system-level goal, the reward for the $m$-th agent at time slot $n$ is defined as the negative value of its current receiver-side AoI:
\begin{equation}
    r_m[n] = -\Delta_m^{\text{Re}}[n].
\end{equation}

This reward formulation directly aligns with the system-wide optimization objective. In particular, maximizing the expected return in the RL framework is equivalent to minimizing the network AoI. In addition, because each agent maintains its own receiver-side AoI counter, the reward signal is fully observable at the local level. This localized feedback allows agents to independently associate their actions with corresponding outcomes, thereby enabling fully decentralized policy optimization without reliance on global coordination.

%
%

\subsection{Decentralized Actor and Centralized Critic Design}
To effectively minimize the system-wide average AoI under partial observability and interference constraints, we adopt a CTDE paradigm. Each agent $m$ is equipped with an actor network $\pi _{{\theta _m}}({{\bf{a}}_m}{\left| \bf{s} \right._m})$, parameterized by $\theta_m$, which maps the local state $\mathbf{s}_m$ to a hybrid action $\mathbf{a}_m$. The hybrid action $\mathbf{a}_m$ consists of a discrete and a continuous component. The discrete branch, $\pi _{{\theta _m}}^{{\rm{disc}}}({{\bf{a}}_m}{\left| \bf{s} \right._m})$, outputs a softmax distribution over $\gamma _m \in \{0, 1\}$, while the continuous branch, $\pi _{{\theta _m}}^{{\rm{cont}}}({{\bf{a}}_m}{\left| \bf{s} \right._m})$, outputs the mean $\mu_m$ of a Gaussian distribution for transmit power, with a learnable and constrained standard deviation $\sigma$.


In particular, the actor's joint policy is expressed as:
\begin{equation}
\pi_{\theta_m}(a_m \!\mid\! \mathbf{s}_m)\! =\! \pi_{\theta_m}^{\text{disc}}(\gamma_m \!\mid\! \mathbf{s}_m) \cdot \pi_{\theta_m}^{\text{cont}}(p_m \! \mid \! \mathbf{s}_m),
\end{equation}
where $\pi^{\text{disc}}$ is modeled by a Categorical distribution over discrete actions and $\pi^{\text{cont}}$ by a Gaussian distribution over continuous power levels.

To facilitate policy learning across all agents, a shared centralized critic $V_\phi(\mathbf{S})$, parameterized by $\phi$, is employed, which takes the full joint state $\mathbf{S} = [\mathbf{s}_1; \dots; \mathbf{s}_M] \in \mathbb{R}^{M \times 7}$ as input. This critic models the value function:
\begin{equation}
    V_\phi(\mathbf{S}) = \left[ V_\phi^{(1)}(\mathbf{S}), \dots, V_\phi^{(M)}(\mathbf{S}) \right] \in \mathbb{R}^M,
\end{equation}
where each output $V_\phi^{(m)}(\mathbf{S})$ corresponds to the estimated return for agent $m$. The critic network in our implementation consists of the following key components:
\begin{itemize}
    \item Feature decomposition: Each agent's local state $\mathbf{s}_m$ is split into GNN feature $f_m$ and the remaining vector $\tilde{\mathbf{s}}_m = \mathbf{s}_m \setminus f_m$, where $\setminus$ denotes element-wise exclusion;
    \item Dual encoding: Separate multi-layer perceptron (MLP) encoders are used for $f_m$ and $\tilde{\mathbf{s}}_m$, producing embeddings $\mathbf{z}_m^{\text{gnn}}$ and $\mathbf{z}_m^{\text{loc}}$, respectively;
    \item Fusion and global encoding: The per-agent embeddings are concatenated as $\mathbf{z}_m = [\mathbf{z}_m^{\text{gnn}} \, \| \, \mathbf{z}_m^{\text{loc}}]$, and all $\mathbf{z}_m$ are flattened and processed through a global MLP to capture inter-agent dependencies;
    \item Value head: The resulting global feature is shared across individual output heads $V_\phi^{(m)}$ for each agent.
\end{itemize}


During training, each agent's policy is optimized using the proximal policy optimization (PPO) algorithm, with the advantage estimates provided by the centralized critic. The total actor loss consists of three components. First, the PPO surrogate loss encourages policy improvement while limiting deviation from the previous policy. For agent $m$, it is given by
\begin{equation}
    \mathcal{L}_{\text{PPO}}^m \!=\! \mathbb{E}_n \left[ \min \left( r_n^m A_n^m,\, \text{clip}(r_n^m, 1 - \epsilon, 1 + \epsilon) A_n^m \right) \right],
\end{equation}
where $r_n^m$ is the ratio of current to previous action probabilities, and $A_n^m$ denotes the advantage function computed via generalized advantage estimation. The hyperparameter $\epsilon$ controls the clipping range to ensure training stability. Second, an entropy regularization term, $\mathcal{L}_{\text{ent}}^m = - \beta \cdot \mathcal{H}(\pi_{\theta_m})$, is incorporated to encourage exploration, where $\mathcal{H}(\pi_{\theta_m})$ denotes the entropy of the hybrid action distribution, encompassing both discrete and continuous components. The coefficient $\beta$ controls the strength of entropy regularization and promotes exploration during training. Empirically, selecting $\beta$ within the range $[0.01, 0.05]$ yields a robust tradeoff between sufficient exploration and stable convergence. Third, an $\ell_2$-norm regularization term, $\mathcal{L}_{\text{reg}}^m = \lambda_{\text{reg}} \cdot \left\| \theta_m \right\|_2^2$, is added to improve generalization and mitigate overfitting, where $\lambda_{\text{reg}}$ regulates the penalty imposed on the magnitude of the policy parameters. Combining these components, the overall loss function used to train the actor network of agent $m$ is given by
\begin{equation}
    \mathcal{L}_{\text{total}}^m = \mathcal{L}_{\text{PPO}}^m + \mathcal{L}_{\text{ent}}^m + \mathcal{L}_{\text{reg}}^m.
\end{equation}

During execution, each agent makes decisions based solely on its local observation $\mathbf{s}_m$ by evaluating its actor network $\pi_{\theta_m}$, without requiring access to the global state or any coordination among agents. 

\subsection{GNN Embedding Optimization}
In addition to providing high-level structural information at the beginning of each episode, the GNN module in the proposed framework is optimized throughout training to enhance its representational quality. Rather than treating the GNN as a fixed feature extractor, we supervise its output embeddings using AoI-related performance signals, enabling the model to generate task-relevant node representations that are aligned with the underlying scheduling utility.

In decentralized multi-agent systems, agents must rely on compact features to approximate complex and partially observable interactions, such as interference coupling, spatial correlation, and long-term reward attribution. Handcrafted metrics or unsupervised GNN pretraining methods often fail to capture the task-specific utility required for effective decision-making. To overcome this limitation, we introduce a supervision mechanism that uses each agent's advantage value, computed through the centralized critic using GAE, as a soft target to guide the learning of GNN representations.

Let $A_m[n]$ denote the advantage of agent $m$ at time slot $n$. After normalizing advantages across time and agents, we compute the mean advantage for each agent over the episode:
\begin{equation}
    \bar{A}_m = \frac{1}{T} \sum_{n=1}^{T} A_m[n].
\end{equation}

To incorporate long-term performance information into the topology-driven GNN embeddings, we exploit the episode-level average advantages as guidance signals. Specifically, we align the pairwise variations of advantages with the pairwise variations of embeddings, for any pair of links. Particularly, if their average advantages are similar, the corresponding embeddings should also be close; otherwise the embeddings should be far away from each other in the case of significant discrepancy of advantage. This leads to the performance-driven metric loss
\begin{equation}\label{GNN_loss}
\begin{aligned}
&{\cal L}_{\rm GNN}=
\frac{1}{|{\cal E}'|}
\sum_{(i,j)\in{\cal E}'}\\
&\left(
\exp\left(-\frac{\|f_i - f_j\|_2^2}{\tau_1}\right)
-\exp \left( -\frac{|\,\bar A_i - \bar A_j\,|}{\tau_2} \right)
 \right)^2,
\end{aligned}
\end{equation}
where ${\cal E}' = \left\{ {\left( {i,j} \right) \in {\cal E}\left| {i \ne j} \right.} \right\}$, $\tau_1$ and $\tau_2 $ are temperature parameters controlling how sensitively embedding differences respond to advantage deviations. By minimizing the discrepancy between embedding-based similarity and advantage-based similarity across link pairs, the proposed GNN loss enforces consistency between the learned representation geometry and long-term performance differences. The resulting pairwise, kernel-based objective preserves the relative performance structure induced by interference and queue dynamics, while remaining insensitive to the absolute scale of the critic outputs, making it well suited for shaping topology-aware representations for MARL.

This GNN update is performed once per episode after computing the full batch of advantage values. This AoI-supervised embedding learning strategy transforms the GNN from a static encoder into a performance-aware feature generator. By anchoring the node features to long-term agent utility, we ensure that each agent receives a condensed summary of the global structure that is directly useful for decision-making. During centralized training, the GNN parameters are learned jointly with the MARL policy. In the distributed implementation, these parameters are fixed, and the GNN is evaluated only once per episode to compute $f_m$ based on the current large-scale fading realization. The proposed MAPPO with GNN algorithm is summarized in Algorithm \ref{algorithm}. Each episode in both training and evaluation corresponds to one large-scale fading interval, during which inter-vehicle distances and path-loss terms remain fixed. Small-scale fading coefficients are re-sampled at every time slot. Across episodes, the large-scale fading parameters are regenerated to expose the MARL agents to diverse vehicular topologies and channel conditions.

\begin{algorithm}[H]
\caption{MAPPO with GNN for V2V Communication}\label{algorithm}
\begin{algorithmic}[1]
\Statex \textbf{Centralized Training}
\State \textbf{Initialize} environment, GNN, actors, critic, optimizers
\State \textbf{Repeat} for each episode:
\State \textbf{Reset} environment and state
\State GNN aggregation: Extract global large-scale information and generate embeddings
\State \textbf{For each time slot:}
\State \quad Generate state: Combine GNN embeddings with local observations
\State \quad Actor network: Select actions
\State \quad Store actions, states, and rewards in memory
\State \quad \quad \textbf{For each agent:}
\State \quad \quad \quad Calculate transmission rate, update buffer, and refresh AoI
\State \quad \quad \quad Compute reward based on AoI
\State \quad \quad \textbf{End For}
\State \textbf{End For}
\State \textbf{If} enough slots:
\State \quad Update Actor and Critic networks
\State \textbf{End If}
\State Update GNN per episode
\State \textbf{End Repeat}
\Statex\rule[1pt]{0.46\textwidth}{0.05em}
\Statex \textbf{Decentralized Execution}
\State \textbf{Load} trained GNN and actor parameters; discard the critic
\State \textbf{Repeat} for each test episode:
\State \textbf{Reset} environment and state
\State GNN aggregation: Compute embeddings once from large-scale information
\State \textbf{For each time slot:}
\State \quad Form local input: Combine cached GNN embedding with local state
\State \quad Actor network: Output packet-dropping decision and transmit power
\State \quad Execute action: Calculate rate, update buffer, and update AoI
\State \textbf{End For}
\State \textbf{End Repeat}
\end{algorithmic}
\end{algorithm}

\section{Simulation Results}\label{s4}
In this section, simulation results are presented to validate the proposed algorithm for V2V AoI optimization. Referencing the work in \cite{18, 14}, the environment simulates a V2V communication network with $M$ V2V links, where the V2V network operates within a rectangular area. Vehicles are dynamically placed across multiple lanes in both horizontal and vertical directions. The large-scale path loss depends on inter-vehicle distances and mobility, whereas the small-scale fading coefficients are independently generated according to a Rayleigh block-fading model. The system bandwidth is set to 1 MHz, and the simulation assumes a noise power spectral density of -174 dBm/Hz, resulting in a system noise power of -114 dBm. The channel coherence time is assumed to be 1 ms, typical for a carrier frequency of 2 GHz and a relative vehicle speed of 36 km/h. Major simulation parameters are listed in Table \ref{parameters}. Unless otherwise specified, the number of packets per batch is set to $u=3$, with a packet arrival probability of 0.8 and each packet having a length of 3000 bits.

\begin{figure}[t]
\begin{center}
\vspace{-3mm}
  \includegraphics[width=3.2in,height=2.4in]{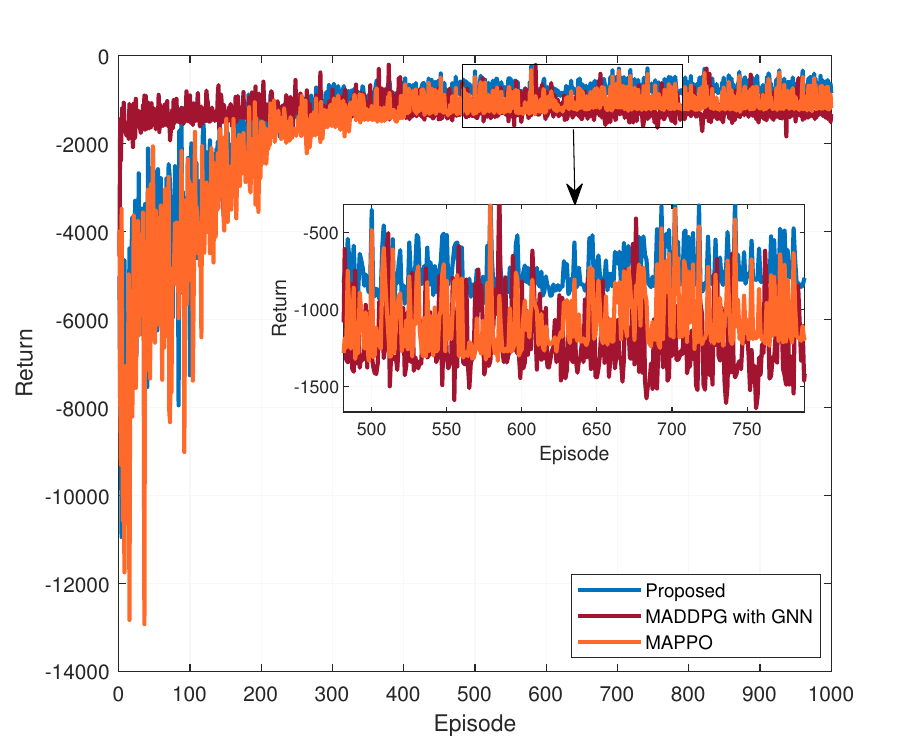}
  \caption{Return for each training episode with increasing iterations.}\label{training_return}
\end{center}
\vspace{-15pt}
\end{figure}

To evaluate the effectiveness of the proposed framework, we compared it against the following baseline strategies. All methods follow the same packet replacement rule in buffer management. When two batches coexist in the buffer and a new batch arrives, the newly arrived batch replaces the existing later-arrived one. The key differences lie in the packet-dropping policy and transmit power control strategy:
\begin{itemize}
\item \textbf{ITLinQ} \cite{33}: A classical interference-aware link scheduling and power control algorithm. In this setup, packet dropping is not allowed; all queued packets are retained for transmission.
\item \textbf{WMMSE} \cite{34}: A weighted minimum mean square error–based algorithm that optimizes transmit power to improve signal quality and reduce interference. It does not support packet dropping and always retains incoming data.
\item \textbf{Random Policy}: Each agent randomly selects both the packet-dropping decision and the transmission power at every time slot.
\item \textbf{AoI-Threshold Policy} \cite{31}: This heuristic strategy adopts a simple rule based on the receiver's AoI. If the AoI exceeds a threshold of 3, the agent drops the current batch and transmits with 70\% of the maximum power. Otherwise, it retains the batch and transmits with 30\% of the maximum power.
\item \textbf{MADDPG with GNN}: A multi-agent deep deterministic policy gradient (MADDPG) \cite{42} framework equipped with the same GNN encoder as the proposed method. The GNN extracts topology-aware embeddings from large-scale fading information, which are incorporated into the decentralized actors, while a centralized critic is trained using global information.
\item \textbf{MAPPO}: A MAPPO-based framework that follows the same CTDE architecture as the proposed method, where the GNN encoder is replaced with a standard MLP operating on local observations only. This baseline serves as an ablation study to isolate and evaluate the impact of the GNN component.
\end{itemize}
\begin{table}[t]
\centering
\caption{Simulation Parameters \cite{18, 14}}\label{parameters}
\label{parameters}
\begin{tabular}{ll}
\toprule
\textbf{Parameter} & \textbf{Value} \\
\midrule
Number of V2V links & 4 \\
Carrier frequency & 2 GHz \\
Vehicle antenna height & 1.5 m \\
Vehicle antenna gain & 3 dBi \\
Vehicle receiver noise figure & 9 dB \\
Vehicle speed & 36 km/h \\
Bandwidth & 1 MHz \\
Noise variance & $-104$ dBm \\
Maximum transmit power & 10 dBm \\
Communication duration & 100 ms \\
Packet length & 3000 bits \\
\bottomrule
\end{tabular}
\end{table}
The training return curves in Fig. \ref{training_return} illustrate the evolution of the cumulative reward during learning for different methods. For the proposed GNN-enhanced MAPPO framework, the return increases steadily with training episodes and gradually stabilizes around episode 400, indicating effective policy improvement and convergence to a near-optimal AoI-aware strategy. The MADDPG with GNN baseline exhibits a faster initial increase in return, reflecting the aggressive updates enabled by deterministic policy gradients. However, its performance saturates at a lower level, suggesting limited long-term optimality despite rapid convergence. In contrast, the MAPPO baseline without the GNN module converges more slowly and attains a lower final return, highlighting the difficulty of learning interference-aware policies without explicit topological feature aggregation. Overall, the proposed method achieves the best balance between convergence speed and final performance, demonstrating that the joint integration of GNN-based topology modeling and MAPPO-based policy optimization is crucial for effective AoI reduction in dynamic V2V networks.

\begin{figure}[t]
\begin{center}
\vspace{-3mm}
  \includegraphics[width=3.2in,height=2.4in]{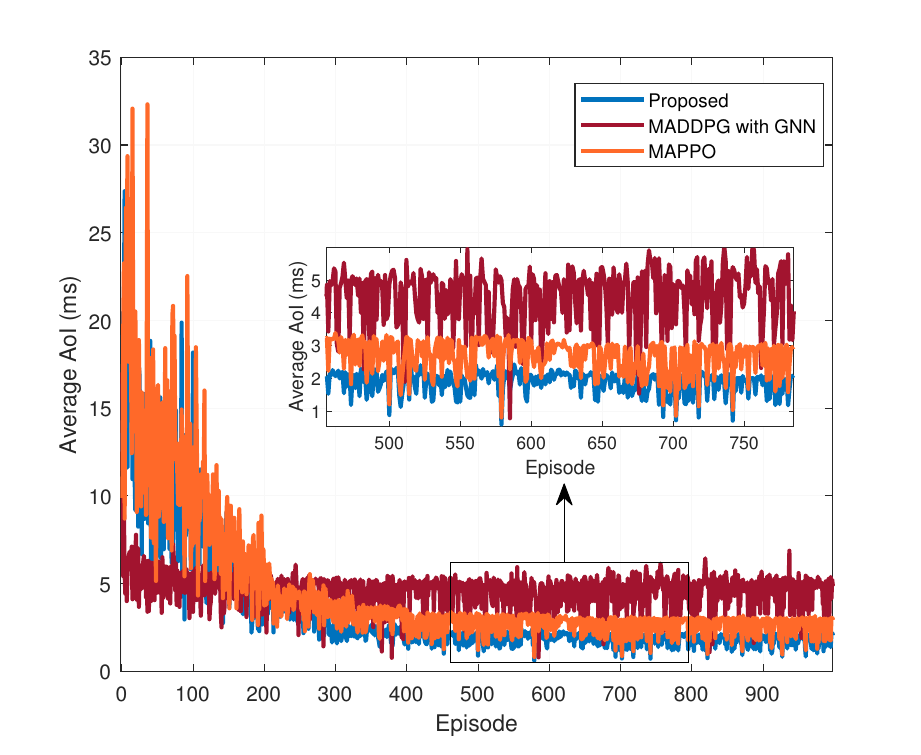}
  \caption{The average AoI for each training episode with increasing iterations.}\label{training_aoi}
\end{center}
\vspace{-15pt}
\end{figure}

The average AoI curve during the training process, as shown in Fig. \ref{training_aoi}, shows the performance of the proposed algorithm across 1000 episodes. Initially, the average AoI value is high, indicating that the system has not yet learned effective strategies for average AoI minimization. As the number of episodes increases, we observe that the AoI for the proposed method decreases rapidly, especially in the first 100 episodes. The proposed method converges to a steady and low AoI value of approximately 1.8 ms, which is close to the duration of a single time slot. This indicates that the learned policy refreshes status information almost every slot, reflecting near-optimal performance for AoI minimization in the V2V communication network. The MADDPG with GNN baseline exhibits a faster initial reduction in AoI, reflecting its rapid policy updates. However, its AoI converges to a higher steady-state value, suggesting suboptimal long-term performance despite fast learning. In contrast, the MAPPO baseline without the GNN module shows a slower AoI reduction and converges to a noticeably higher AoI.

\begin{figure}[t]
\begin{center}
\vspace{-3mm}
  \includegraphics[width=3.2in,height=2.3in]{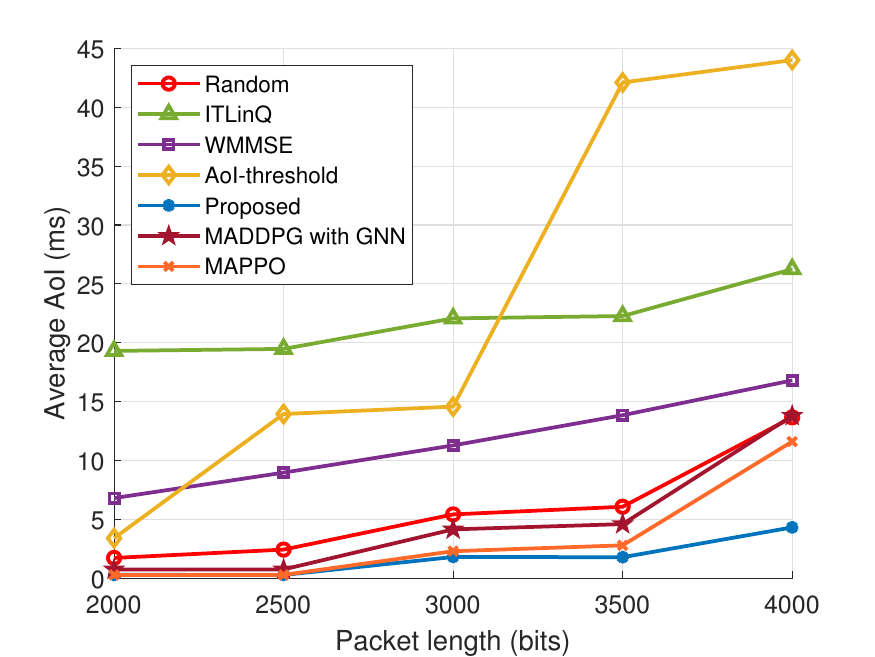}
  \caption{The impact of packet length on average AoI.}\label{packet_length}
\end{center}
\vspace{-10pt}
\end{figure}

Fig. \ref{packet_length} illustrates the average AoI achieved by different algorithms as a function of packet length, ranging from 2000 to 4000 bits. As expected, the AoI increases with packet length for all methods, reflecting the higher transmission latency incurred by larger data payloads. However, the rate of increase differs markedly across algorithms. The proposed method consistently attains the lowest AoI over the entire range of packet lengths, exhibiting only a mild degradation as packet size grows, which demonstrates strong robustness to increasing transmission demand. Both learning-based baselines, namely MADDPG with GNN and MAPPO, also outperform traditional rate-centric methods as packet length increases, indicating the advantage of learning-based AoI-aware control. The MADDPG with GNN baseline benefits from topology-aware representations and shows improved scalability compared with non-learning approaches. However, its AoI remains consistently higher than that of the proposed method, suggesting that deterministic policy gradients are less effective in capturing long-term AoI trade-offs. The MAPPO baseline without the GNN module exhibits inferior performance to both the proposed method and MADDPG with GNN, particularly at larger packet sizes, highlighting the importance of explicit interference modeling through graph-based feature aggregation. In contrast, ITLinQ and WMMSE exhibit poor AoI performance despite their effectiveness in maximizing rate-oriented metrics. These algorithms prioritize links with high SINR, which leads to prolonged service delays for links under unfavorable channel conditions. As a result, packets from disadvantaged links accumulate in the buffer, significantly degrading the average AoI. Interestingly, the random policy achieves non-negligible performance, which can be attributed to its implicit compatibility with the system’s resource constraints. From a queue management perspective, the limited buffer size of two batches enables random dropping to frequently eliminate stale packets, passively preserving data freshness. From a power control perspective, although random power allocation does not adapt to channel conditions, stochastic fluctuations may occasionally align with favorable fading realizations, enabling partial packet delivery. By contrast, the fixed AoI-threshold strategy rapidly reaches the preset AoI limit as packet length increases, causing frequent packet drops. This behavior severely limits successful batch delivery and results in a pronounced AoI increase. Overall, these results confirm that while learning-based approaches scale better with packet length than conventional methods, the proposed GNN-enhanced MAPPO framework achieves the most robust AoI performance by jointly leveraging topology-aware representations and stochastic policy optimization.

\begin{figure}[t]
\begin{center}
\vspace{-3mm}
  \includegraphics[width=3.2in,height=2.3in]{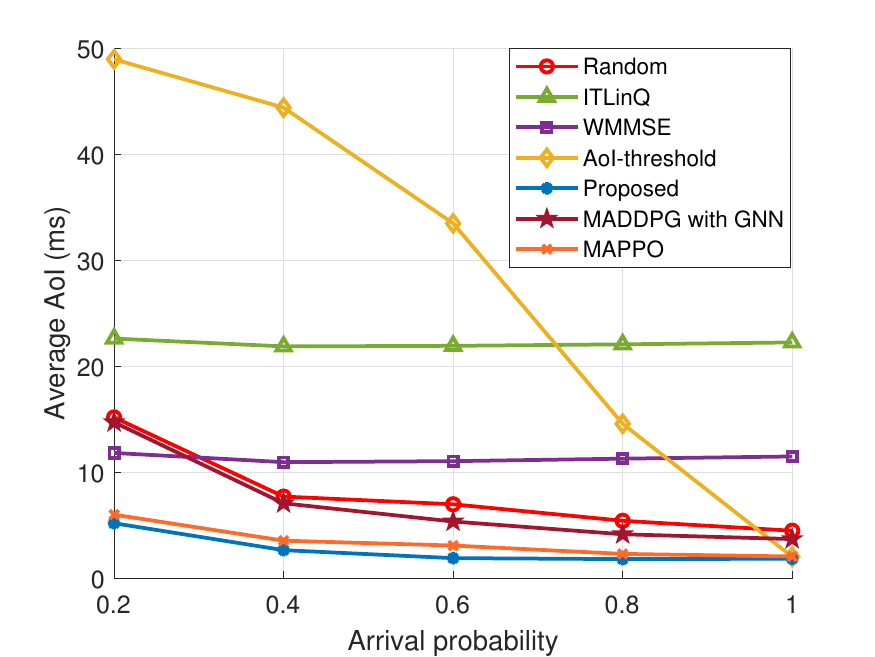}
  \caption{The impact of packet arrival probability on average AoI.}\label{arrival_probability}
\end{center}
\vspace{-10pt}
\end{figure}

Fig. \ref{arrival_probability} provides valuable insights into how the packet arrival probability influences the AoI for various algorithms. As the packet arrival probability increases, the AoI generally decreases for most of the methods. This is expected because as the arrival probability increases, the system has more opportunities to refresh the information, thus reducing the AoI at the receiver end. The learning-based approaches, including MADDPG with GNN, MAPPO, and the proposed method, consistently outperform traditional heuristics across the entire range of arrival probabilities. Among them, the proposed framework achieves the lowest AoI and exhibits the most stable performance. Despite the random nature of packet dropping, a higher arrival probability helps reduce the AoI since more packets are available for transmission, allowing random choices to occasionally result in better packet delivery. It is worth noting that the monotonic decreasing trend in Fig. \ref{arrival_probability} differs from the classical ``U-shaped" AoI behavior observed in single-packet queuing models \cite{7}. In our batch-based setting, outdated batches can be actively dropped, preventing persistent queue buildup even at the highest tested arrival rates. As a result, the system remains within the low-congestion regime where increasing the arrival rate improves freshness without inducing queue instability.

\begin{figure}[t]
\begin{center}
\vspace{-3mm}
  \includegraphics[width=3.2in,height=2.3in]{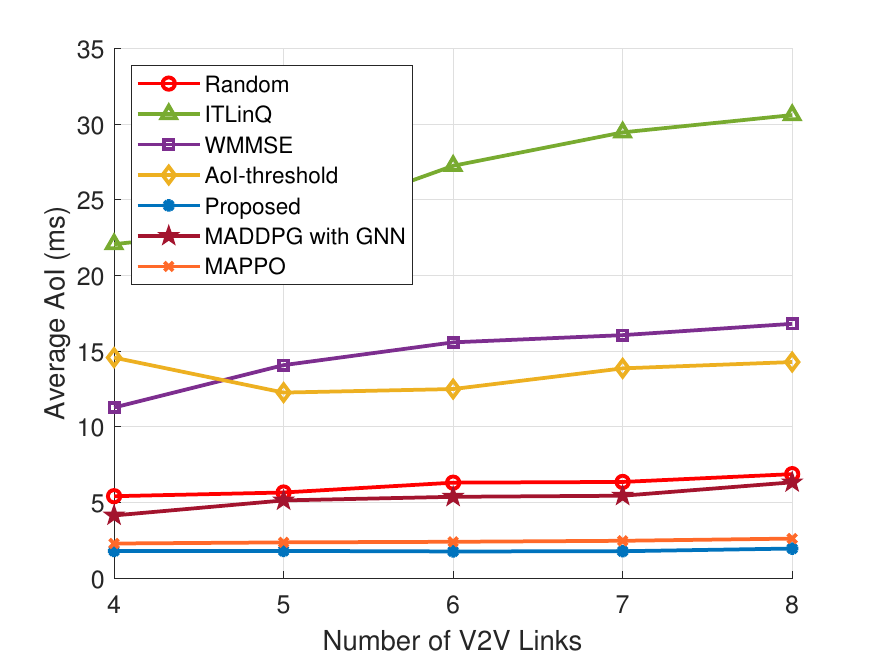}
  \caption{Comparison of average AoI under different numbers of V2V links.}\label{veh_number}
\end{center}
\vspace{-10pt}
\end{figure}

\begin{figure*}[t]
\vspace{-15pt}
\begin{center}
\subfloat[Discrete action over time slot for each V2V link.]{
\includegraphics[width=3.5in,height=2.2in]{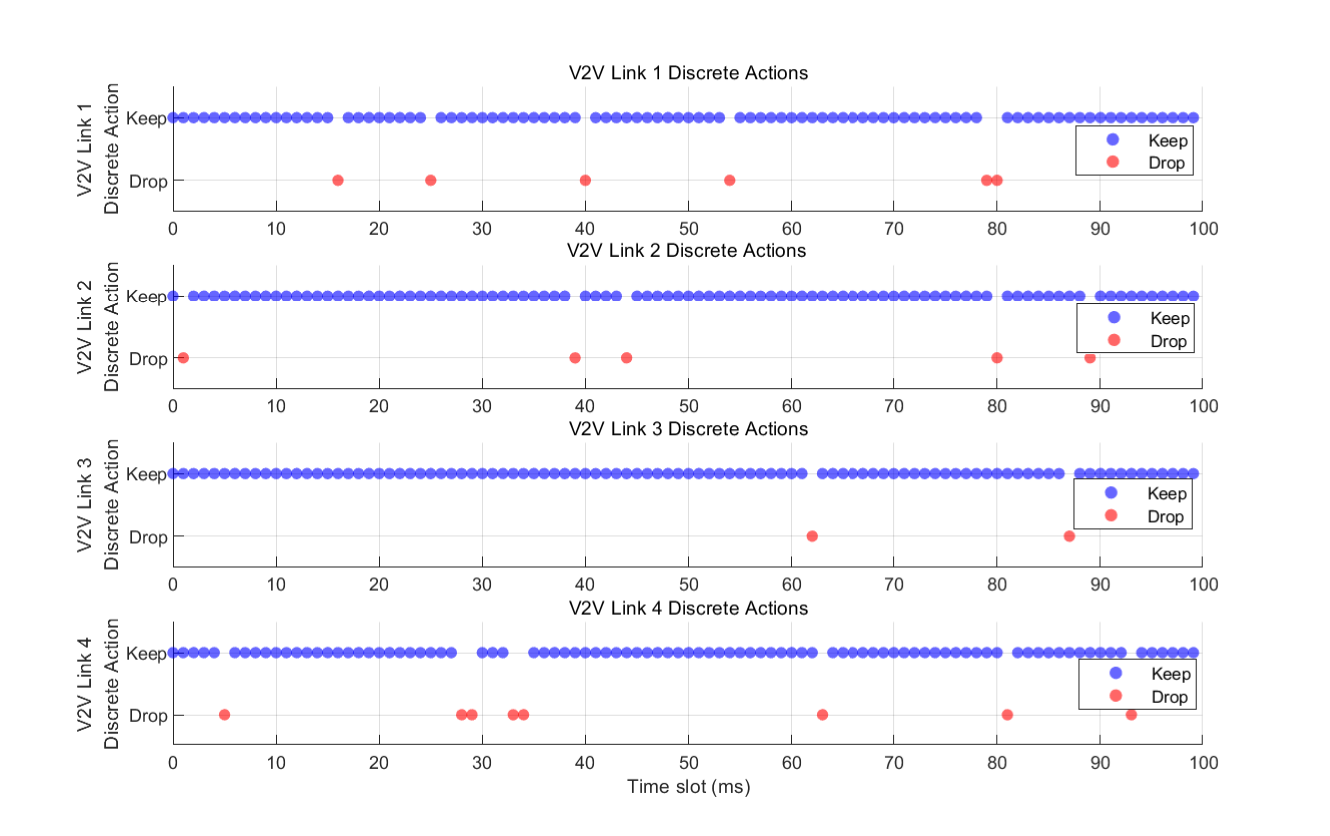}\label{discrete_actions}
}
\subfloat[Continuous action over time slot for each V2V link.]{
\includegraphics[width=3.5in,height=2.2in]{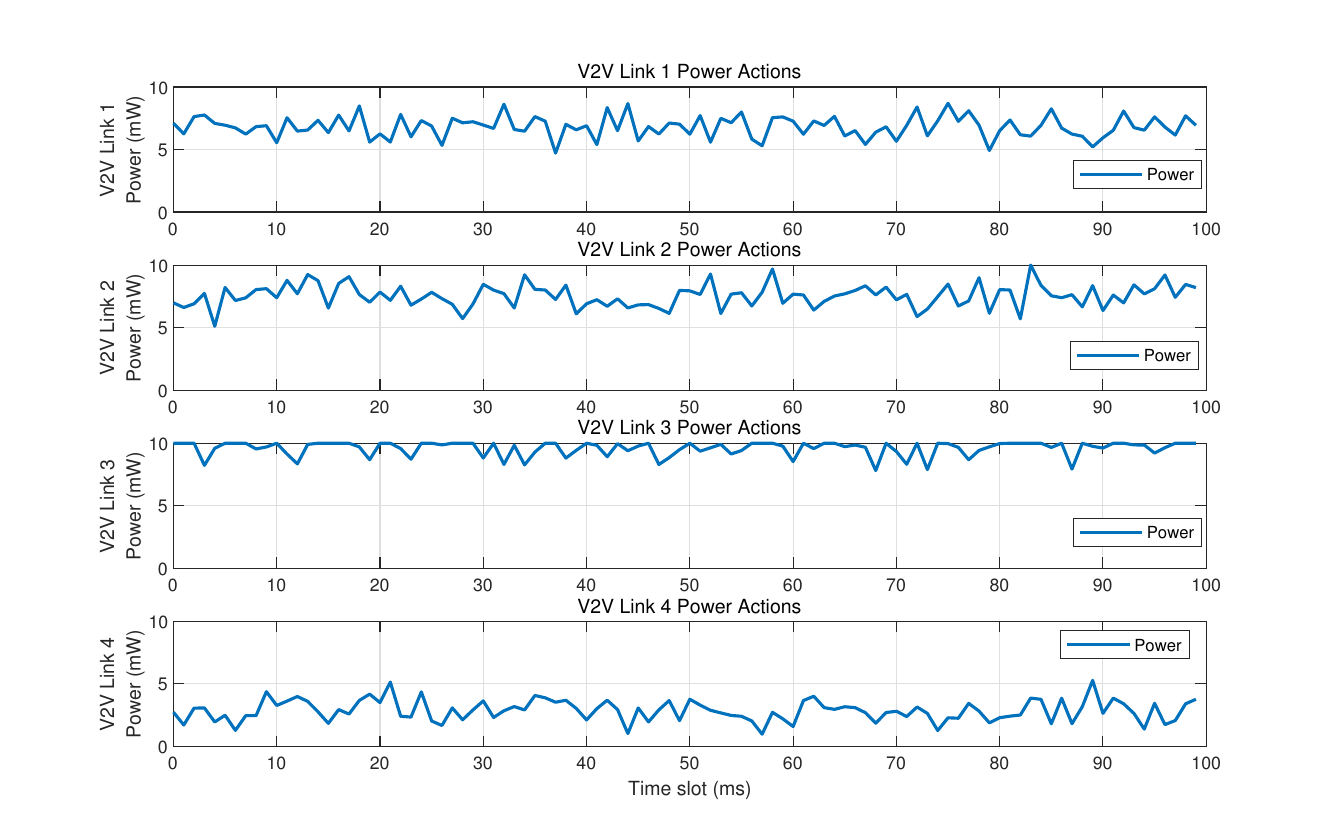}\label{power_actions}
}
\caption{Discrete (a) and continuous (b) actions selected by the proposed method across all V2V links.}
\label{actions}
\end{center}
\vspace{-15pt}
\end{figure*}

\begin{figure*}[t]
\vspace{-15pt}
\begin{center}
\subfloat[AoI evolution over time slot under the proposed method.]{
\includegraphics[width=3.5in,height=2.2in]{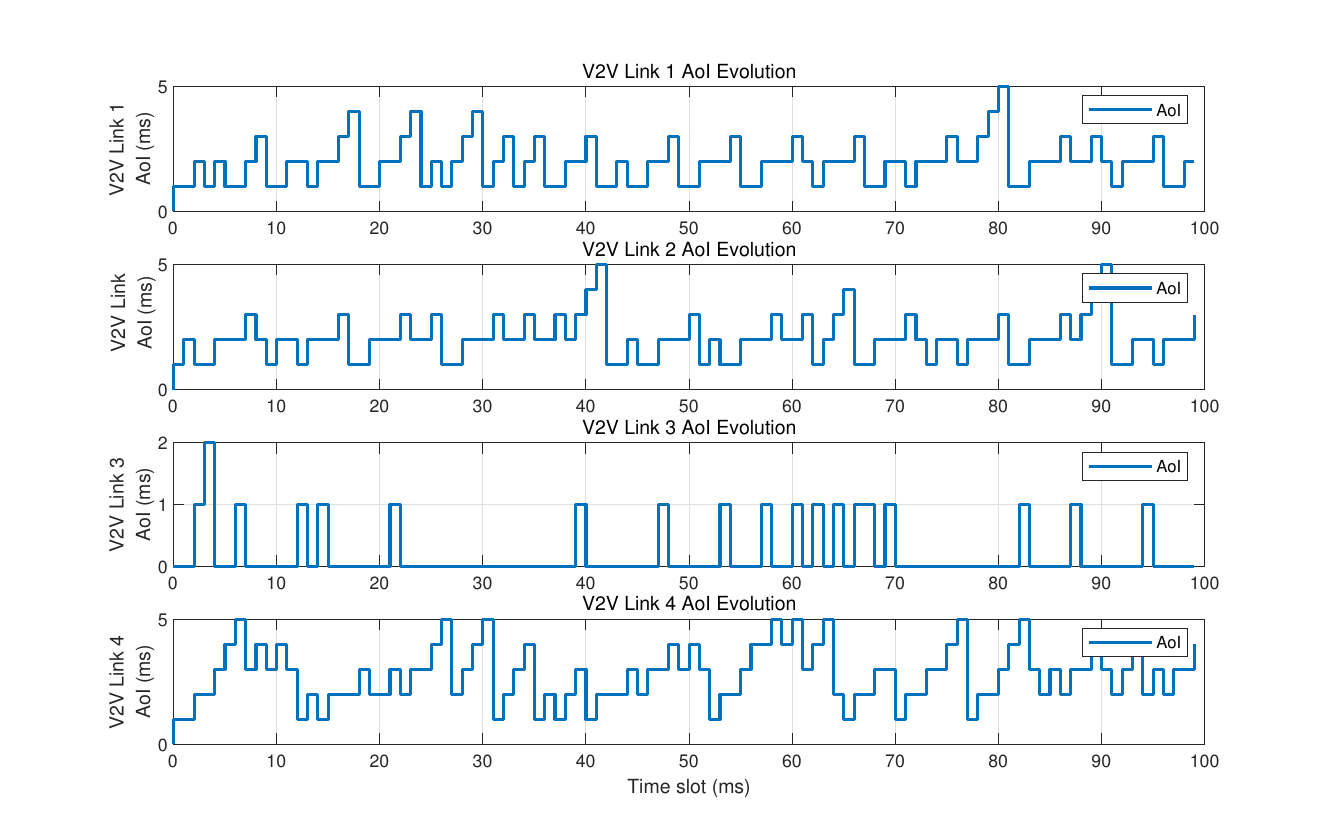}
}
\subfloat[AoI evolution over time slot under the random policy.]{
\includegraphics[width=3.5in,height=2.2in]{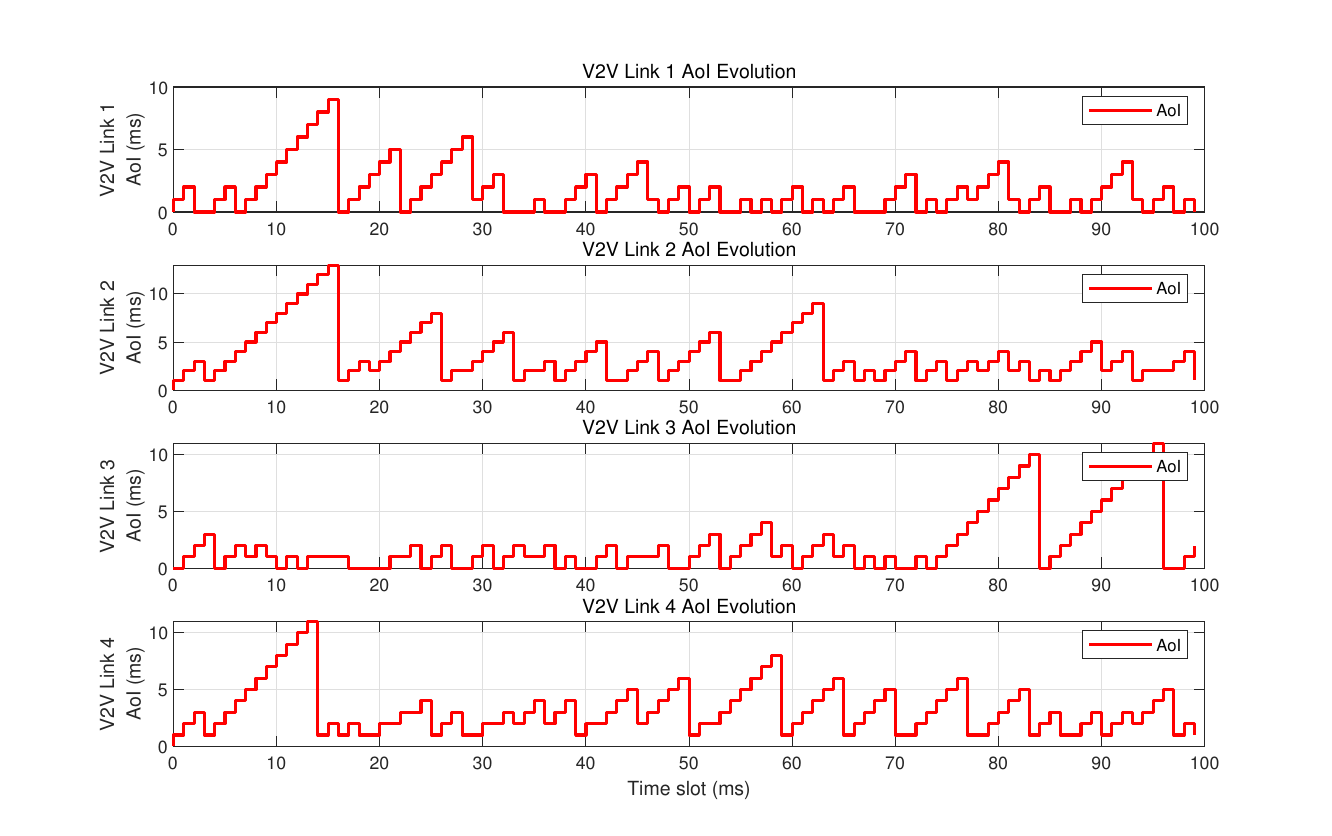}
}
\caption{Comparison of per-link AoI evolution under (a) the proposed method and (b) the random method.}
\label{aoi_evolution}
\end{center}
\vspace{-15pt}
\end{figure*}

Fig. \ref{veh_number} illustrates the average AoI performance of different algorithms as the number of V2V links increases from 4 to 8. As the network becomes denser, all methods experience an increase in average AoI due to intensified competition for limited radio resources and stronger mutual interference. Despite this challenge, the proposed method consistently achieves the lowest AoI across all network sizes, demonstrating superior scalability and coordination efficiency under dense vehicular deployments. The learning-based baselines, including MADDPG with GNN and MAPPO, also outperform traditional heuristic approaches, indicating the benefit of adaptive, data-driven control in handling increased network density. In contrast, heuristic methods such as ITLinQ and WMMSE, which lack queue-awareness and inter-agent coordination, show limited adaptability as network density increases. Their rate-centric designs lead to inefficient service differentiation under strong interference, resulting in rapidly deteriorating AoI performance. Overall, these results confirm that the proposed GNN-enhanced MARL framework scales more effectively with network size by jointly capturing interference structure and long-term AoI dynamics.

\begin{table*}[t]
\centering
\caption{Average AoI under different packet lengths with packet arrival probability of 0.8}\label{t2}
\begin{tabular}{ccccccc}
\toprule
\textbf{Packet Length (bits)} & 1500 & 2000 & 2500 & 3000 & 3500 & 4000 \\
\midrule
\textbf{Average AoI (ms)} & 0.369 & 0.471 & 1.040 & 1.797 & 2.374 & 4.247 \\
\bottomrule
\end{tabular}
\end{table*}

\begin{table*}[t]
\centering
\caption{Average AoI under different arrival probabilities with packet length of 3000 bits}\label{t3}
\begin{tabular}{cccccc}
\toprule
\textbf{Arrival Probability} & 0.2 & 0.4 & 0.6 & 0.8 & 1.0 \\
\midrule
\textbf{Average AoI (ms)} & 5.514 & 2.832 & 1.992 & 1.797 & 1.886 \\
\bottomrule
\end{tabular}
\vspace{-10pt}
\end{table*}

Fig. \ref{actions} shows the V2V link actions based on the proposed method. Fig. \ref{actions}(a) illustrates the discrete actions of each V2V link, while Fig. \ref{actions}(b) presents the power allocation actions. It can be observed that at certain times, the V2V links choose to drop packets, indicating that the proposed method effectively makes decisions based on environmental feedback. In V2V links 1 to 4, power allocation fluctuates according to the demands of each link. Through these two figures, we can see that the MAPPO algorithm, when executed, makes adaptive discrete actions and power adjustment decisions based on the varying needs of the links. As time progresses, these actions and power allocations reflect the resource optimization strategy of the algorithm in the V2V communication network. This dynamic adjustment mechanism can effectively enhance system performance in complex communication environments.

Fig. \ref{aoi_evolution} presents the evolution of the AoI over time for four V2V links under two different strategies: the proposed method, as shown in Fig. \ref{aoi_evolution}(a), and the Random method, as shown in Fig. \ref{aoi_evolution}(b). The MAPPO strategy clearly outperforms the Random strategy in maintaining a low and stable AoI across all V2V links. While the AoI increases over time under both strategies, the dynamic adjustments of the proposed method, based on system feedback, allow it to keep AoI under control. In contrast, the Random strategy leads to significant fluctuations and higher AoI values. The proposed method exhibits a more efficient use of available resources, ensuring that the information remains fresh for longer periods, even in a dynamic environment. This highlights the effectiveness of the proposed method in optimizing the freshness of information, compared to the Random strategy, which lacks adaptive decision-making and leads to outdated information.

The performance of the proposed method was evaluated under varying packet lengths, as shown in Table \ref{t2}. The training environment involved five vehicles with a packet arrival probability of 0.8 and a packet length of 3000 bits. For testing, the packet length was varied from 1500 bits to 4000 bits to assess its impact on AoI. As expected, the average AoI increases with packet size, primarily due to the longer transmission times required for larger packets. Specifically, smaller packet sizes (1500 to 2000 bits) result in relatively low AoI, indicating efficient status updates. However, as the packet length grows (2500 to 4000 bits), AoI rises significantly, particularly at 3000 and 4000 bits, where AoI reaches 1.797 and 4.247, respectively. This suggests that larger packets introduce greater delays in status updates, leading to higher AoI. Despite this, our method remains robust, maintaining relatively low AoI values even with larger packets, highlighting the proposed method's ability to balance communication overhead and timely status updates effectively. This adaptability confirms the model's resilience in real-world scenarios, where packet size can fluctuate.

The performance of the proposed method was evaluated under varying packet arrival probabilities, as shown in Table \ref{t3}. The training environment parameters are as described in Table \ref{t2}, with the packet arrival probability during testing varied from 0.2 to 1.0. At a low arrival probability of 0.2, the average AoI is significantly higher, as fewer packets arrive, leading to more frequent outdated information. However, as the packet arrival probability increases, the AoI decreases due to more frequent status updates. The average AoI stabilizes at 1.886 when the arrival probability reaches 1.0, demonstrating that higher packet arrival rates facilitate more timely information updates and effectively reduce AoI. Notably, the optimal AoI performance is achieved at an arrival probability of 0.8, suggesting that this balance between packet arrival rate and system processing is most effective for AoI reduction. These results confirm that the system benefits from more frequent updates, with AoI decreasing as the packet arrival probability increases. Our method demonstrates strong adaptability to changes in packet arrival rates, maintaining minimized AoI across varying conditions. This further highlights the effectiveness of our method.

\section{conclusion}\label{s5}
This paper presented a GNN-enhanced multi-agent reinforcement learning framework to jointly optimize queue management and resource allocation for average AoI minimization in V2V communication networks. By modeling each vehicle’s status as a batch of multiple interdependent packets, the proposed method effectively captured the realistic characteristics of vehicular sensing data and addressed the critical trade-off between timely information delivery and limited transmission resources. We integrated a topology-aware GNN into a CTDE paradigm based on MAPPO and further aligned the learned embeddings with long-term agent utility to improve representation quality and decision relevance. Through this design, the proposed framework provided a scalable and adaptive solution for AoI-sensitive queue management and resource control in dynamic vehicular environments. It effectively learned to coordinate hybrid discrete-continuous actions under partial observability and time-varying interference. The sum-AoI objective used in this work focuses on system-wide efficiency and does not explicitly address fairness across links. Future extensions may incorporate fairness-aware objectives (e.g., max-AoI, AoI variance, or weighted utilities) or fairness-driven reward shaping in MARL. In addition, the framework can be extended to support event-driven status updates and status-sampling-based packet scheduling, enabling more practical and fine-grained AoI optimization under real-world sensing and communication constraints.


\bibliographystyle{IEEEtran}
\bibliography{myref}

\end{document}